\documentclass[aps,showpacs,eqsecnum,floatfix]{revtex4}
\usepackage{amsbsy}

\usepackage[dvips]{graphicx,color}

\begin{document}

\title{A New Approach to the 3D Faddeev Equation for Three-Body
Scattering}

\author{Ch.~Elster$^{(a)}$}

\author{W.~Gl\"ockle$^{(b)}$}

\author{H.~Wita{\l}a$^{(c)}$}

\affiliation{(a)
Institute of Nuclear and Particle Physics,  and
Department of Physics and Astronomy,  Ohio University,
Athens, OH 45701,
USA}
\affiliation{(b)
Institute for Theoretical Physics II, Ruhr-University Bochum,
D-44780 Bochum, Germany}

\affiliation{(c) 
M. Smoluchowski Institute of Physics, Jagiellonian
 University, PL-30059 Krak\'ow, Poland}

\date{\today}

\vspace{10mm}

\begin{abstract}
A novel approach to solve the Faddeev equation for three-body scattering at
arbitrary energies is proposed. This approach disentangles the
complicated singularity structure of the free three-nucleon propagator
leading to the moving and logarithmic singularities in standard
treatments. 
The Faddeev equation is formulated 
in momentum space and directly solved in terms of momentum vectors
without employing a partial wave decomposition.  In its simplest form
the Faddeev equation for identical bosons, which we are using , is an
integral equation in five variables, magnitudes of
relative momenta and angles. 
The singularities of the free propagator and the deuteron propagator
are now both simple poles in two different momentum variables, 
and thus
can both be integrated with standard techniques.

\end{abstract}
\vspace{10mm}

\pacs{21.45.-v,25.10.+s}

\maketitle


\section{Introduction}

In 1960 L.D.~Faddeev formulated his mathematically rigorous scattering
theory for three particles by proposing a set of three coupled integral
equations, which do have a unique solution ~\cite{faddeev}. In the first 
numerical
realizations of this approach, separable interactions were introduced,
reducing the three-body equations to a set of one-dimensional coupled
integral equations, whose numerical solutions was feasible at the
time~\cite{mitra,lovelace}. Despite a simplification 
due to the choice of 
the two-body interaction, the standard formulation of the momentum space
Faddeev equations in the continuum contains for the free three-body
propagator a complicated singularity structure within the integral
kernel~\cite{schmiedz,redish,wgbook}. These complications arise because the
position of the propagator cut does not only depend on the total energy
of the system but also on external momentum variables on a grid (thus
the term moving singularities). In addition branch points occur
leading to logarithmic singularities. 

It is possible to carry out the integration along a path in the
complex plane, thus avoiding to directly deal with these singularities,
but indirectly imposing conditions on the analytical properties of 
the two-body force.
This method of contour deformation
was introduced for separable potentials by Hetherington and
Schick~\cite{HSchick} and perfected by Cahill and Sloan~\cite{CalSloan}. 
The need to use realistic forces, which are predominantly local, led to 
methods integrating the singularities on the real momentum
axis~\cite{KloetTjon}. It took until the 1980's until the Faddeev
equations were solved in the continuum with a realistic nucleon-nucleon
(NN) force as input~\cite{FBS,Witaa:1989zz}. 

During the last two decades calculations of nucleon-deuteron scattering
experienced large improvements and refinements. It is fair to say that
below about the pion production  projectile energy the momentum space Faddeev
equations for three-nucleon scattering can now be solved with high
accuracy for the most modern two- and three-nucleon forces. A summary
of these achievements can be found in
Refs.~\cite{wgphysrep,wgarticle,witala1,Epelbaum:2005pn,Deltuva:2005cc}. 
The approach described there is based on using angular
momentum eigenstates for the two- and three-body systems. This partial
wave decomposition replaces the
continuous angle variables by discrete orbital angular momentum quantum
numbers, and thus reduces the
number of continuous variables to be discretized in a numerical
treatment to two.  For low projectile energies the procedure of considering
orbital angular momentum components appears physically justified due
to arguments related to the centrifugal barrier and the short range of
the nuclear force. However, when considering three nucleon scattering
at higher energies, it appears natural to avoid a partial wave
representation completely and work directly with vector variables. 
Only recently exact Faddeev calculations for three-body scattering
in the intermediate energy regime
became available. The formulation 
and numerical realization based on vector variables
 for the nonrelativistic Faddeev
equations~\cite{Liu:2004tv,Elster:2008yt}
as well as fully Poincar{\'e} invariant
ones~\cite{Lin:2007kg,Lin:2008sy} have been carried out for
scalar interactions up to projectile energies of 2~GeV.

Despite the technical sophistication with which the Faddeev equations in
the continuum are solved today, the treatment of the singularity structure
of the free three-nucleon propagator experienced only minor modifications 
from the original suggestion~\cite{Witaa:1989zz}, e.g. in
Ref.~\cite{Liu:2004tv} the logarithmic singularities are integrated
semi-analytically with splines in contrast to the earlier subtraction
techniques. However, it would be most desirable to have a kernel
without any logarithmic singularities. In Ref.~\cite{Witala:2008my}
a solution to this long lasting
technical challenge is proposed and successfully carried out
 in the context of the partial wave decomposed Faddeev equations.  
It is the purpose of this paper to examine this suggestion in the
context of a three-dimensional treatment of the Faddeev equation and
introduce a kernel in which only simple poles in one variable occur. 
Those poles can than be integrated by standard subtraction techniques.

In Section II we briefly revisit the form of the nonrelativistic
Faddeev equation used in previous work~\cite{Liu:2004tv} to allow
an easy comparison of the differences in our 
new approach. In Section III
we describe in detail the simplification of the singularity structure 
of the free three-body propagator, and in Section IV we complete 
the calculation
with the remaining angular integrations and connect
to previous work in calculating the operators for elastic and
breakup scattering. We conclude in Section V.

\section{The Faddeev Equation for Three Identical Bosons}

There are various presentations of three-body scattering in the Faddeev
scheme~\cite{faddeev}
 presented in the literature~\cite{alt,wgphysrep,wgbook}. We consider
here the Faddeev equation for identical particles in the form
\begin{equation}
T|\phi\rangle = tP|\phi\rangle + tPG_0 T |\phi\rangle .
\label{eq:2.1}
\end{equation}
The driving term of this integral equation consists of a two-body
t-matrix $t$, the sum $P$ of a cyclic and anticyclic permutation of three
identical particles, and the initial state $|\phi\rangle = |\varphi_d {\bf
q_0} \rangle$, composed of a two-body bound state and the
momentum eigenstate of the projectile particle. The kernel of
Eq.~(\ref{eq:2.1}) contains the free three-body propagator, $G_0 = (E-H_0
+i \epsilon)^{-1}$, where $E$ is the total energy in the
center-of-momentum (c.m.) frame.
The operator $T$ determines both the full breakup amplitude
\begin{equation}
U_0=(1+P)T
\label{eq:2.1b}
\end{equation}
and the amplitude for elastic scattering
\begin{equation}
U=PG_0^{-1}+PT ~.
\label{eq:2.1c}
\end{equation}
For the explicit solution of
Eq.~(\ref{eq:2.1}) the standard Jacobi momenta {\bf p}, the relative
momentum in the subsystem, and {\bf q}, the relative momentum of the
spectator to the subsystem are introduced. The momentum states are
normalized according to $\langle {\bf p}'{\bf q}' \vert {\bf p}{\bf q}
\rangle = \delta^3({\bf p}' -{\bf p}) \; \delta^3 ({\bf q}' -{\bf q})$.
Projecting Eq.~(\ref{eq:2.1}) onto these basis states leads to
\begin{eqnarray}
T({\bf p},{\bf q};{\bf q_0})& =& T_0({\bf p},{\bf q};{\bf q_0})  \cr
 &+& \int d^3 p' d^3 p'' d^3q''
 t({\bf p},{\bf p}';\varepsilon) \langle {\bf p}'
{\bf q} | P | {\bf p}''{\bf q~}'' \rangle
  \frac{1}{E + i\epsilon - E''} T({\bf p}'',{\bf q}'';{\bf q_0}).
\label{eq:2.2}
\end{eqnarray}
Here we abbreviate $\langle {\bf p}{\bf q}|T|\varphi_d {\bf q_0}\rangle \equiv
T ({\bf p},{\bf q};{\bf q_0})$ and the driving term as $\langle {\bf p}{\bf
q}|tP|\varphi_d {\bf q_0}\rangle \equiv T_0 ({\bf p},{\bf q};{\bf q_0})$.
Under the integral we
 take advantage of the fact that the two-body $t$-matrix only depends on
the relative momenta {\bf p} and {\bf p}' of the subsystem, and
the thus the  $t$-matrix is evaluated
at the energy $\varepsilon = E -\frac{3}{4m}q^2$.
The energy $E''$ of the free three-body propagator is given by
\begin{equation}
E''  =   \frac{1}{m}( p''^2 + \frac{3}{4} q''^2) ~.
\label{eq:2.3}
\end{equation}
The permutation operator is explicitly given as
\begin{equation}
\langle {\bf p}'{\bf q} |P| {\bf p}'' {\bf q}''\rangle =
  \delta({\bf p}'+ \boldsymbol{\pi}_1) \; \delta( {\bf p}'' -\boldsymbol{\pi}_2)
 +  \delta({\bf p}' -\boldsymbol{\pi}_1)\; \delta ({\bf p}'' +
\boldsymbol{\pi}_2),
\label{eq:2.4}
\end{equation}
with the `shifted' momenta
\begin{eqnarray}
\boldsymbol{\pi}_1 & = &  \frac{1}{2} {\bf q} + {\bf q}'' \nonumber \\
\boldsymbol{\pi}_2 & = &   {\bf q} + \frac{1}{2}  {\bf q}'' ~.
\label{eq:2.5}
\end{eqnarray}
Inserting Eqs.~(\ref{eq:2.3})-(\ref{eq:2.5}) into Eq.~(\ref{eq:2.2})
leads to
\begin{eqnarray}
T({\bf p},{\bf q},{\bf q_0}) & = &  T_0({\bf p},{\bf q},{\bf q_0}) +
  \int d^3 p' d^3 p'' d^3 q'' \; t( {\bf p},{\bf p}~';\varepsilon)  \cr
& & \Big[\delta({\bf p}'+\boldsymbol{\pi}_1) \; \delta({\bf p}'' -\boldsymbol{\pi}_2)
  +\delta({\bf p}' -\boldsymbol{\pi}_1) \; \delta ({\bf p}~''+\boldsymbol{\pi}_2)
\Big]  \cr
& &   \frac{1}{E + i\epsilon - E''}  \; T({\bf p}'',{\bf q}'',{\bf q_0}).
\label{eq:2.6}
\end{eqnarray}
The direct evaluation of the matrix elements of the
permutation operator gives
\begin{eqnarray}
T({\bf p},{\bf q},{\bf q_0}) & = &  T_0({\bf p},{\bf q},{\bf q_0}) +
 \int  d^3 q'' \cr
& & \Big[ t( {\bf p},-\boldsymbol{\pi}_1;\varepsilon) \frac{1}{E + i\epsilon -
E''} \; T(\boldsymbol{\pi}_2,{\bf q}'',{\bf q_0}) \cr
& & + t( {\bf p},\boldsymbol{\pi}_1;\varepsilon)  \frac{1}{E + i\epsilon - E''}
 \; T(-\boldsymbol{\pi}_2,{\bf q}'',{\bf q_0}) \Big].
\label{eq:2.7}
\end{eqnarray}
Defining a symmetrized t-matrix
\begin{equation}
t_s({\bf p},\boldsymbol{\pi}_1;\varepsilon) =
t({\bf p},\boldsymbol{\pi}_1;\varepsilon)
+ t({\bf p},-\boldsymbol{\pi}_1;\varepsilon)
\label{eq:2.8}
\end{equation}
and realizing that for identical bosons
$T(\boldsymbol{\pi}_2,{\bf q}'';{\bf q_0}) =
T(-\boldsymbol{\pi}_2,{\bf q}'';{\bf q_0})$
one arrives at the expression
\begin{eqnarray}
T({\bf p},{\bf q},{\bf q_0})  &=&   T_0({\bf p},{\bf q},{\bf q_0}) \cr
&+&  \int  d^3 q'' t_s({\bf p},\boldsymbol{\pi}_1;\varepsilon) \frac{1}{E +
i\epsilon - \frac{1}{m}(q^2 +q''^2 +{\bf q}\cdot {\bf q}'')}
T(\boldsymbol{\pi}_2,{\bf q}'',{\bf q_0}),
\label{eq:2.9}
\end{eqnarray}
which is the starting point for the numerical calculations presented in
Ref.~\cite{Liu:2004tv}. The free propagator in Eq.~(\ref{eq:2.9})
clearly displays the difficulties inherent in numerically
solving the three-body scattering problem in this form. The propagator
depends on the magnitude of $q''$  and through the scalar product ${\bf
q}\cdot {\bf q}''$ on the angle between ${\bf q}''$ and a fixed axis
given by ${\bf q}$. The integration over $d^3 q''$ leads to
singularities with respect to that angle for each fixed value
of $q''$ and $q$. These singularities are integrable, but lead to
logarithmic singularities in the variable $q''$. However, despite being
integrable, they pose  numerical challenges. For scattering calculations
in a three-dimensional approach these challenges were met in
Ref.~\cite{Liu:2004tv,Elster:2008yt} for the nonrelativistic Faddeev
equation and in
Ref.~\cite{Lin:2007kg,Lin:2008sy} for Poincar\'e invariant Faddeev
equation. Nonetheless, three-body scattering calculations would be
less challenging, if these singularities in the angles could be extricated
from the ones in the momenta. A suggestion for accomplishing this task
is made in the  section III, following ~\cite{Witala:2008my}.

\section{A New Look at the Singularity Structure of the Free Three-Body
Propagator}

As demonstrated in the previous section, using the vector variables in the evaluation
of the matrix elements of the permutation operator of Eq.~(\ref{eq:2.4}) leads to the
representation of the free three-body propagator of Eq.~(\ref{eq:2.9}), containing both
integration variables $q''$ and ${\hat {\bf q}}\cdot {\hat {\bf q}}''$. If we want to
disentangle those two variables, we should start by separating the angular piece
of the delta functions from the piece containing magnitudes of momenta and
write
\begin{eqnarray}
\delta({\bf p}' + \boldsymbol{\pi}_1)\; \delta ({\bf p}'' -\boldsymbol{\pi}_2)
&=&
\frac{\delta(p' -\pi_1)}{p'^2} \frac{\delta (p''-\pi_2)}{p''^2}
\delta({\hat {\bf p}}' + {\hat {\boldsymbol{\pi}}}_1 ) \delta ({\hat {\bf p}}'' -
{\hat {\boldsymbol{\pi}}}_2)\cr
\delta({\bf p}' - \boldsymbol{\pi}_1)\; \delta ({\bf p}'' +\boldsymbol{\pi}_2)
&=&
\frac{\delta(p' -\pi_1)}{p'^2} \frac{\delta (p''-\pi_2)}{p''^2}
\delta({\hat {\bf p}}' -{\hat {\boldsymbol{\pi}}}_1 ) \delta ({\hat {\bf p}}'' +
{\hat {\boldsymbol{\pi}}}_2)
\label{eq:3.1}
\end{eqnarray}
This leads to
\begin{eqnarray}
T({\bf p},{\bf q},{\bf q_0}) &=& T_0({\bf p},{\bf q},{\bf q_0}) +
  \int dp' p'^2 d{\hat {\bf p}}' \; dp'' p''^2 d{\hat {\bf p}}'' \;
   d^3 q'' t({\bf p},{\bf p}',\varepsilon)  \cr
& & \Big[ \delta( {\hat {\bf p}}' + {\hat {\boldsymbol{\pi}}}_1) \;
\delta ({\hat {\bf p}}'' - {\hat {\boldsymbol{\pi}}}_2) +
\delta({\hat {\bf p}}' - {\hat {\boldsymbol{\pi}}}_1 ) \;
\delta ({\hat {\bf p}}'' + {\hat {\boldsymbol{\pi}}}_2) \Big] \cr
& & \frac{\delta(p' -\pi_1)}{p'^2} \frac{\delta (p'' -\pi_2)}{p''^2}
 \frac{1}{E + i\epsilon - E''}\; T({\bf p}'',{\bf q}'',{\bf q_0}) \cr
& = &   T_0({\bf p},{\bf q},{\bf q_0}) + \int d^3 q'' \; dp' \; dp''
  \delta(p' -\pi_1) \; \delta (p''-\pi_2)\cr
& & \frac{1}{E + i\epsilon - \frac{1}{m}(p''^2 +\frac{3}{4}q''^2)}
(t_s( {\bf p}, p'{\hat {\boldsymbol{\pi}}}_1; \varepsilon)
T(p''{\hat {\boldsymbol{\pi}}}_2, {\bf q''}, {\bf q_0}).
\label{eq:3.2}
\end{eqnarray}
We arrived at the last expression by integrating over the angles and
taking into account the symmetrized two-body t-matrix
from Eq.~(\ref{eq:2.8}). 
The magnitudes of the two vectors $\boldsymbol{\pi}_1$ and $\boldsymbol{\pi}_2$
are given by
\begin{eqnarray}
|\boldsymbol{\pi}_1| &=& \sqrt{\frac{1}{4}q^2 +q''^2 + qq''x''} \cr
|\boldsymbol{\pi}_2| &=& \sqrt{q^2 + \frac{1}{4} q''^2 + qq''x''}
\label{eq:3.3}
\end{eqnarray}
where $x''={\hat {\bf q}}\cdot {\hat {\bf q}}''$. The above relations can 
be used
to rewrite the two delta functions of Eq.~(\ref{eq:3.2}) in a form better 
suited
for our further considerations. We rewrite
\begin{equation}
\delta(p'-\pi_1) = \frac{2p'}{qq''} \; \delta(x''-x_0) \; \Theta(1-|x_0|)
\label{eq:3.4a}
\end{equation}
where
\begin{equation}
x_0 = \frac{1}{qq''} \left( p'^2 -\frac{1}{4}q^2 -q''^2 \right) =
    \frac{1}{qq''} \left( p''^2 -\frac{1}{4}q''^2 -q^2 \right)
\label{eq:3.4b}
\end{equation}
and
\begin{equation}
\delta(p'-\pi_2) =\delta \left( p'' -\sqrt{p'^2 +\frac{3}{4} q^2 -\frac{3}{4} q''^2} \right)
 \; \Theta \left( p'' -\sqrt{p'^2 +\frac{3}{4} q^2 -\frac{3}{4} q''^2} \right) 
~.
\label{eq:3.4c}
\end{equation}
Inserting these expressions into Eq.~(\ref{eq:3.2}) gives
 \begin{eqnarray}
T({\bf p},{\bf q},{\bf q_0}) &=& T_0({\bf p},{\bf q},{\bf q_0})  \cr
&+& \int d^3 q'' \; dp' \; dp''
 \frac{2p'}{qq''} \; \delta(x'' -x_0)\; \Theta(1-|x_0|)  \cr
& &\delta \left( p''-\sqrt{p'^2+\frac{3}{4}q^2-\frac{3}{4}q''^2} \right) \;
 \Theta \left(p'^2+\frac{3}{4}q^2-\frac{3}{4}q''^2 \right) \cr
& & \frac{1}{E + i\epsilon - \frac{1}{m}(p''^2 +\frac{3}{4}q''^2)}
\; t_s( {\bf p}, p'{\hat {\boldsymbol{\pi}}}_1; \varepsilon) \;
T(p''{\hat {\boldsymbol{\pi}}}_2, {\bf q''}, {\bf q_0}).
\label{eq:3.5}
\end{eqnarray}

\begin{figure}
\begin{center}
 \includegraphics[width=10cm]{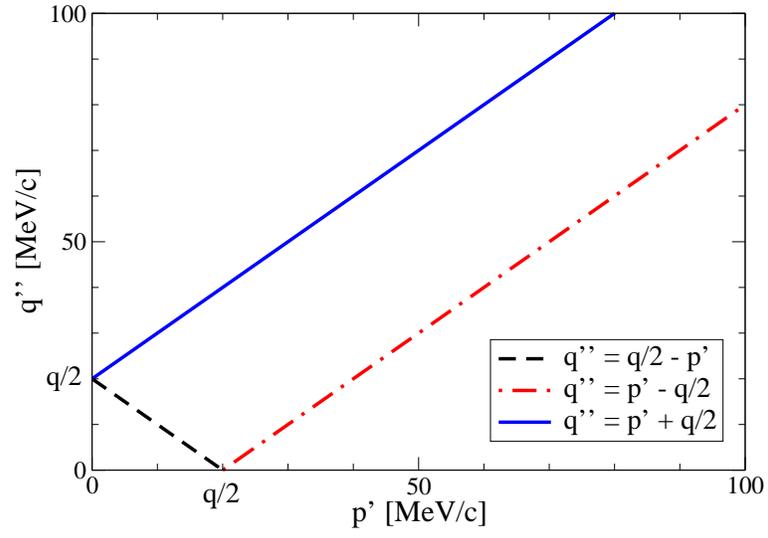}
\end{center}
\caption{(Color online) The domain for the integration over the momenta
$p'$ and $q''$ as function of the external momentum $q$. 
Here $q$~=~40~MeV/c is chosen. The area of integration is the rectangle 
enclosed
by the three lines.
\label{fig1}}
\end{figure}

Before we continue, we remember that the underlying two-body force supports one bound state
with energy $E_d$. This means, that $t_s({\bf p},{\bf p}';z)$ has a pole at $z=E_d$.
Because the transition operator $T$ of Eq.~(\ref{eq:3.5}) is needed for all values of
${\bf q}$, one will encounter this pole of $t_s$. Extracting the residue explicitly
by defining
\begin{equation}
t_s({\bf p},{\bf p}';z) \equiv \frac{{\hat t}_s ({\bf p},{\bf p}';z)}{z-E_d}
\label{eq:3.6}
\end{equation}
and similarly for $T$ and $T_0$ one can rewrite Eq.~(\ref{eq:3.5}) as
 \begin{eqnarray}
{\hat T}({\bf p},{\bf q},{\bf q_0}) &=& {\hat T_0}({\bf p},{\bf q},{\bf q_0})  \cr
&+& \int d^3 q'' \; dp' \; dp''
 \frac{2p'}{qq''} \; \delta(x'' -x_0)\; \Theta(1-|x_0|)  \cr
& &\delta \left( p''-\sqrt{p'^2+\frac{3}{4}q^2-\frac{3}{4}q''^2} \right) \;
 \Theta \left(p'^2+\frac{3}{4}q^2-\frac{3}{4}q''^2 \right) \cr
& & \frac{1}{E + i\epsilon - \frac{1}{m}(p''^2 +\frac{3}{4}q''^2)}
 {\hat t}_s( {\bf p}, p'{\hat {\boldsymbol{\pi}}}_1; \varepsilon) \;
\frac{{\hat T}(p''{\hat {\boldsymbol{\pi}}}_2, {\bf q''}, {\bf q_0})}
{E+i\epsilon -E_d -\frac{3}{4m}q''^2}.
\label{eq:3.7}
\end{eqnarray}
Now we carry out the integration in $p''$ and arrive at
\begin{eqnarray}
{\hat T}({\bf p},{\bf q},{\bf q_0}) &=& {\hat T_0}({\bf p},{\bf q},{\bf q_0})  \cr
&+& \frac{2}{q} \int d{\hat {\bf q}}'' \; d q'' \; dp'
  \; \delta(x'' -x_0)\; \Theta(1-|x_0|) \;
 \Theta \left(p'^2+\frac{3}{4}q^2-\frac{3}{4}q''^2 \right) \cr
& & \frac{p'q''}{E + i\epsilon - \frac{1}{m}(p'^2 +\frac{3}{4}q^2)}
 {\hat t}_s( {\bf p}, p'{\hat {\boldsymbol{\pi}}}_1; \varepsilon) \;
\frac{{\hat T}(p''{\hat {\boldsymbol{\pi}}}_2, {\bf q''}, {\bf q_0})}
{E+i\epsilon -E_d -\frac{3}{4m}q''^2},
\label{eq:3.8}
\end{eqnarray}
where $p'' \equiv \sqrt{p'^2+\frac{3}{4}q^2 -\frac{3}{4}q''^2}$. 
The two theta-functions in
Eq.~(\ref{eq:3.8}) restrict the integration in the magnitudes of 
$q''$ and $p'$ into an area
whose size depends on the magnitude of the spectator momentum $q$.
An example of this area is indicated in Fig.~\ref{fig1} for the
choice of $q$~=~40~MeV/c.

Next we consider the product of propagators in Eq.~(\ref{eq:3.8}) and 
separate this
product as in Ref.~\cite{Witala:2008my}
\begin{eqnarray}
& &\frac{1}{E + i \epsilon - \frac{1}{m}(p'^2 + \frac{3}{4} q^2)}
\frac{1}{E + i\epsilon - E_d -\frac{3}{4m} q''^2}  = \cr
& &  \left[ \frac{1}{E + i \epsilon - \frac{1}{m}( p'^2 + \frac{3}{4} q^2)}
 - \frac{1}{E + i\epsilon - E_d -\frac{3}{4m} q''^2} \right]
 \frac{1}{ - E_d - \frac{3}{4m} q''^2 + \frac{1}{m}( p'^2 + \frac{3}{4} q^2)}.
\label{eq:3.9}
\end{eqnarray}
The new denominator function,
\begin{equation}
{\bar G}(q,q'',p') =
\frac{1}{ - E_d - \frac{3}{4m} q''^2 + \frac{1}{m}( p'^2 + \frac{3}{4} q^2)}
\label{eq:3.10}
\end{equation}
can not become singular inside  the integration domain $p' - q''$. 
Using the relation from Eq.~(\ref{eq:3.4c}) we see that
\begin{equation}
{\bar G}(q,q'',p') = \frac{1}{- E_d  + \frac{1}{m} p''^2}=
\frac{1}{ | E_d|+ \frac{1}{m} p''^2} > 0.
\label{eq:3.11}
\end{equation}
With this separation of propagators the three-body transition amplitude from
Eq.~(\ref{eq:3.8}) consists now of two integrals and can be written as
\begin{eqnarray}
\hat T({\bf p},{\bf q},{\bf q_0}) &=& \hat T_0({\bf p},{\bf q},{\bf q_0}) \cr
&+&  \frac{2}{q} \int_0^{\infty} dp' p'
\frac{1}{E + i \epsilon - \frac{1}{m}( p'^2 + \frac{3}{4} q^2)}
\int_{|q/2 - p'|}^{q/2 + p'} dq'' q'' \; {\bar G}(q,q'',p')  \cr
& & \int d{\hat {\bf q}}'' \; \delta(x''-x_0)
{\hat t}_s( {\bf p}, p'{\hat {\boldsymbol{\pi}}}_1; \varepsilon)
 \; {\hat T}(p''{\hat {\boldsymbol{\pi}}}_2, {\bf  q''}, {\bf  q_0})  \cr
&-& \frac{2}{q}\int_0^{\infty} dq'' q'' \frac{1}{E + i\epsilon - E_d -\frac{3}{4m} q''^2}
 \int_{|q/2-q''|}^{q/2+q''} dp' p' \; {\bar G}(q,q'',p') \cr
& &  \int d{\hat {\bf q}}'' \delta(x''-x_0)
{\hat t}_s({\bf p},p'{\hat {\boldsymbol{\pi}}}_1; \varepsilon) \;
{\hat T}(p''{\hat {\boldsymbol{\pi}}}_2, {\bf  q''}, {\bf  q_0}) ~.
\label{eq:3.12}
\end{eqnarray}
In the second part of the kernel we changed the sequence of
integrations over $p'$ and $q''$.

This new form of the kernel now exhibits only simple poles in the
$p'$ and $q''$ integration, and the angle integration does {\bf not}
contain any singularity any more. As a remark, the simultaneous
occurence of poles in the angle and momentum integration when solving
Eq.~(\ref{eq:2.9}) leads to the logarithmic singularities, which are not
present in Eq.~(\ref{eq:3.12}).
The poles of the free propagator occur
in the variable $p'$ and the deuteron pole in
the variable $q''$.


\section{The Angle Integration}
Since we ignore spin and isospin dependencies, the matrix element
$ T({\bf p},{\bf q},{\bf q_0})$ is a scalar function of the variables
${\bf p}$ and ${\bf q}$ for a given projectile momentum ${\bf q_0}$.
As  was shown in Ref.~\cite{Liu:2004tv}, one needs 5 variables to uniquely
specify the geometry of those three vectors. For the clarity of presentation we
repeat some of the arguments here. Having in mind that
with three vectors one can span two planes, i.e. the
$\mathbf{p}$-$\mathbf{q}_{0}$-plane and
the $\mathbf{q}$-$\mathbf{q}_{0}$-plane, a natural choice of independent
variables is
\begin{equation}
p=|{\mathbf{p}}|,\  q=|{\mathbf{q}}|,\
x_{p}=\hat{{\mathbf{p}}}\cdot\hat{{\mathbf{q}}}_{0},\
x_{q}=\hat{{\mathbf{q}}}\cdot\hat{{\mathbf{q}}}_{0},\
x^{q_{0}}_{pq}=
\widehat{({\mathbf{q}}_{0}\times{\mathbf{q}})}\cdot\widehat{({\mathbf{q}}_{0}\times{\mathbf{p}})}.
\label{variables}
\end{equation}
The last variable, $x^{q_{0}}_{pq}$, is the angle between the two normal
vectors of the $\mathbf{p}$-$\mathbf{q}_{0}$-plane and the
$\mathbf{q}$-$\mathbf{q}_{0}$-plane.
It should further be pointed out, that the angle between the vectors
${\bf p}$ and ${\bf q}$, $y_{pq}={\hat {\bf p}}\cdot {\hat {\bf q}}$ is not an
independent variable, but can by related to the ones given above as
\begin{equation}
y_{pq}=x_{p}x_{q}+\sqrt{1-x^{2}_{p}}\sqrt{1-x^{2}_{q}}\
x^{q_{0}}_{pq}. \label{eq:4.2}
\end{equation}
For the special case where
$\hat{\mathbf{q}}_{0}$ is parallel to the $z$-axis ($q_{0}$-system) one can
write
\begin{eqnarray}
y_{pq}=x_{p}x_{q}+\sqrt{1-x^{2}_{p}}\sqrt{1-x^{2}_{q}}\cos\varphi_{pq},
\label{eq:4.3}
\end{eqnarray}
where the angle $\varphi_{pq}$ is the difference of the azimuthal angles of
$\hat{\mathbf{p}}$ and $\hat{\mathbf{q}}$ related to the specified z-axis.

The $\delta$-function in the angle integration of Eq.~(\ref{eq:3.12}),
$\delta(x''-x_0)=\delta({\hat {\bf q}}\cdot {\hat {\bf q}''} -x_0)$ suggests
to choose for the ${\hat {\bf q}}''$ integration
the z-axis parallel to the vector ${\bf q}$ ($q$-system).
The angle dependence of
the two-body t-matrix is then explicitly given as
\begin{equation}
{\hat t}_s ({\bf p},p'{\hat {\boldsymbol{\pi}}}_1;E(q)) \equiv
{\hat t}_s (p,p',{\hat {\bf p}}\cdot {\hat {\boldsymbol{\pi}}}_1; \varepsilon),
\label{eq:4.4}
\end{equation}
where
\begin{equation}
{\hat {\bf p}}\cdot {\hat {\boldsymbol{\pi}}}_1 = \frac{\frac{1}{2}q y_{pq} + y_{pq''}}
{\sqrt{\frac{1}{4}q^2+q''^2+qq''x''}}
\label{eq:4.5}
\end{equation}
and
\begin{equation}
y_{pq''} = {\hat {\bf p}}\cdot {\hat {\bf q}}'' =
y_{pq} x'' +\sqrt{1-y_{pq}^2} \; \sqrt{1 -x''^2} \; \cos(\varphi_p -\varphi'').
\label{eq:4.6}
\end{equation}
The angle $\varphi_p$ is the azimuthal angle of ${\hat {\bf p}}$ in the
$q$-system. As mentioned before, $x''={\hat {\bf q}}\cdot {\hat {\bf q}}''$ and
$q'' =|{\bf q''}|$.

\noindent
The angle dependence of the three-body transition amplitude is more intricate
and can be written as
\begin{equation}
{\hat T}(p''{\hat {\boldsymbol{\pi}}}_2, {\bf  q''}, {\bf  q_0}) \equiv
{\hat T} (p'',x_{\pi_2},x^{q_0}_{\pi_2 y_{q_0 q''}}, y_{q_0 q''},q'';q_0).
\label{eq:4.7}
\end{equation}
Explicitly these angles are given as
\begin{eqnarray}
{\hat {\boldsymbol{\pi}}}_2 \cdot {\bf q_0} &\equiv& x_{\pi_2} =
 \frac{q x_q +\frac{1}{2}q''y_{q_0 q''}}{\sqrt{q^2+\frac{1}{4}q''^2 + qq''x''}}
\nonumber \\
{\hat {\bf q}}'' \cdot {\bf q_0} & \equiv&  y_{q_0 q''} =
  x_q x'' +\sqrt{1-x_q^2}\; \sqrt{1-x''^2} \; \cos (\varphi_{q_0}-\varphi'')
\nonumber \\
x^{q_0}_{\pi_2 y_{q_0 q''}} & \equiv&
\frac{{\hat \pi}_2\cdot {\hat q}'' - ({\hat \pi}_2\cdot {\hat q}_0)({\hat
q}'' \cdot {\hat q}_0)}{\sqrt{1-({\hat \pi}_2\cdot {\hat q}_0)^2}\;
\sqrt{1-({\hat q}'' \cdot {\hat q}_0)^2}}
=\frac{ \frac{qx''-\frac{1}{2}q''}{\sqrt{q^2+\frac{1}{4}q''^2
+qq''x''}}-x_{\pi_2} y_{q_0 q''} }
{\sqrt{1-x_{\pi_2}^2}\; \sqrt{1-y_{q_0 q''}^2}}.
\label{eq:4.8}
\end{eqnarray}
Like $\varphi_p$ in Eq.~(\ref{eq:4.6}) the angle $\varphi_{q_0}$ is the
azimuthal angle of ${\hat {\bf q_0}}$ in the $q$-system. It was shown in
Ref.~\cite{Liu:2004tv} that because of the $\varphi''$ integration, only the
knowledge of $\cos(\varphi_{q_0} - \varphi_p)$ is required. This difference
can be explicitly represented as
\begin{equation}
\cos ( \varphi_{q_0} - \varphi_p) = \frac{ \hat q_0 \cdot \hat p - ( \hat q
\cdot \hat q_0) ( \hat q \cdot \hat p)}{ \sqrt{1- ( \hat q \cdot \hat
q_0)^2 } \sqrt{1 - ( \hat p \cdot \hat q)^2}} = \frac{x_p - y_{pq}
x_q}{\sqrt{1 - y_{pq}^2} \sqrt{1- x_q^2}} ~.
\end{equation}

\noindent
Since only
difference of the angles enters, one can choose  $\varphi_{q_0}$
arbitrarily, e.g. $\varphi_{q_0}=0$.
Furthermore, $\cos \varphi_p$ and $\sin \varphi_p$ required in
Eq.~(\ref{eq:4.6}) are then also uniquely given~\cite{Liu:2004tv}.

With these preparations we are ready to carry out the angle integration
$\int d{\hat {\bf q}}'' = \int_{-1}^{1}dx'' \int_0^{2\pi} d\varphi''$ of
Eq.~(\ref{eq:3.12}) explicitly. The $x''$ integration is fixed by the
$\delta$-function in terms of $x_0 = x_0(q,p',q'')$ from Eq.~(\ref{eq:3.4b}),
leaving only an integration over $\varphi''$. Explicitly, the variables of
Eqs.~(\ref{eq:4.6}) and (\ref{eq:4.8}) need only be evaluated at a fixed
$x''=x_0(q,p',q'')$. Thus, the explicit representation for the transition
amplitude ${\hat T}$ reads
\begin{eqnarray}
\hat{T}(p,x_{p}, x^{q_{0}}_{pq}, x_{q}, q; q_0) &=&
 \hat{T}_0(p,x_{p}, x^{q_{0}}_{pq}, x_{q}, q; q_0)  \label{eq:4.9} \\
&+& \frac{2}{q} \int_0^{\infty} dp' p'
\frac{1}{E + i \epsilon - \frac{1}{m}( p'^2 + \frac{3}{4} q^2)}
\int_{|q/2 - p'|}^{q/2 + p'} dq'' q'' \; {\bar G}(q,q'',p')  \cr
& & \int_0^{2\pi} d\varphi'' \; {\hat t}_s \left(p,p',\frac{\frac{1}{2}q y_{pq} +
y_{pq''}(x_0)}{\sqrt{\frac{1}{2}q^2+q''^2+qq''x_0}};\varepsilon \right) \cr
& &{\hat T} \left( p'', \frac{q x_q+\frac{1}{2}q''
y_{q_0q''}(x_0)}{\sqrt{q^2+\frac{1}{4}q''^2+qq_0x_0}}, x^{q_0}_{\pi_2
y_{q_0 q''}}(x_0), y_{q_0q''}(x_0), q'';q_0 \right) \cr
&-& \frac{2}{q}\int_0^{\infty} dq'' q'' \frac{1}{E + i\epsilon - E_d
-\frac{3}{4m} q''^2}
 \int_{|q/2-q''|}^{q/2+q''} dp' p' \; {\bar G}(q,q'',p') \cr
& & \int_0^{2\pi} d\varphi'' \; {\hat t}_s \left(p,p',\frac{\frac{1}{2}q y_{pq} +
y_{pq''}(x_0)}{\sqrt{\frac{1}{2}q^2+q''^2+qq''x_0}}; \varepsilon \right) \cr
& &{\hat T} \left( p'', \frac{q x_q+\frac{1}{2}q''
y_{q_0q''}(x_0)}{\sqrt{q^2+\frac{1}{4}q''^2+qq_0x_0}}, x^{q_0}_{\pi_2
y_{q_0 q''}}(x_0), y_{q_0q''}(x_0), q'';q_0 \right), \nonumber
\end{eqnarray}
where $p''=\sqrt{p'^2+\frac{3}{4}q^2-\frac{3}{4}q''^2}$ is fixed.

The only remaining detail is to provide an explicit expression for the Born term,
\begin{equation}
T({\bf p},{\bf q},{\bf q_0}) \equiv \langle {\bf p}{\bf q}|tP|\varphi_d {\bf
q_0}\rangle,
\label{eq:4.10}
\end{equation}
where $\varphi_d$ stands for the deuteron bound state. Projecting on Jacobi
momenta and evaluating the permutation operator leads to
\begin{equation}
T({\bf p},{\bf q},{\bf q_0})=\varphi_d
\left( {\mathbf{q}}+\frac{1}{2}{\mathbf{q}}_{0}\right)
t_{s}\left({\mathbf{p}},\frac{1}{2}{\mathbf{q}}+{\mathbf{q}}_{0}, \varepsilon \right).
\label{eq:4.11}
\end{equation}
Using the invariant variables of Eq.~(\ref{variables}) one arrives at
\begin{eqnarray}
 \hat{T}_0(p,x_{p}, x^{q_{0}}_{pq}, x_{q}, q; q_0)
&=&\varphi_{d}\left(\sqrt{q^{2}+\frac{1}{4}q^{2}_{0}+qq_{0}x_{q}}\right)
 \label{eq:4.12}  \\
&\times&\hat{t}_{s} \left( p,
\sqrt{\frac{1}{4}q^{2}+q^{2}_{0}+qq_{0}x_{q}},
\frac{\frac{1}{2}qy_{pq}+q_{0}x_{p}}{\sqrt{\frac{1}{4}q^{2}+q^{2}_{0}+qq_{0}x_{q}}};
\varepsilon \right). \nonumber
\end{eqnarray}
The transition amplitude of Eq.~(\ref{eq:4.9}) together with the Born term
given above has as far as the angle integration is concerned a structure very
similar to the one given in Eq.~(2.19) of Ref.~\cite{Liu:2004tv}. Even in the
case where logarithmic singularities have to be integrated, they  are
independent of the angle $\varphi''$. Thus, in the numerical realization
demonstrated in Ref.~\cite{Liu:2004tv} the  $\varphi''$-integration of the
kernel is carried out for each fixed value $x''$ and $q''$ on their
respective grids. Then the singularity structure depending only on the
$x''$ and $q''$ variables is explicitly dealt with. In Eq.~(\ref{eq:4.9})
the $\varphi''$-integration needs to be carried out for each fixed
value $q''$ and $p'$ given on their respective grids. These grids also
fix $x_0(q,p',q'')$. Since the functional dependence on $\varphi''$ is the
same in both integrals over $\varphi''$ as is the area of integration in
the $p'$-$q''$-plane, the integral over $\varphi''$ in the kernel needs
to be evaluated only once if both integrals of Eq.~(\ref{eq:4.9}) are
calculated on the same $p'$-$q''$-grid. Under these conditions the numerical
effort as far as the angle integration is concerned is similar to the one
of Ref.~\cite{Liu:2004tv}. 

Once the transition operator $T({\bf p},{\bf q},{\bf q_0})$ is explicitly
calculated as function of the 5 independent variables, the amplitude for
elastic scattering is obtained by calculating the matrix elements of the
operator $U$ given in Eq.~(\ref{eq:2.1b}) as
\begin{eqnarray}
\langle{\mathbf{q}}\varphi_{d}|U|{\mathbf{q}}_{0}\varphi_{d}\rangle
&=&2\varphi_{d}\left(\frac{1}{2}{\mathbf{q}}+{\mathbf{q}}_{0}\right)
\left(E-\frac{1}{m}(q^{2}+{\mathbf{q}}\cdot{\mathbf{q}}_{0}+q^{2}_{0})\right)
\varphi_{d}\left({\mathbf{q}}+\frac{1}{2}{\mathbf{q}}_{0}\right)  \nonumber \\
&+&2 \int
d^{3}q''\varphi_{d}\left(\frac{1}{2}{\mathbf{q}}+{\mathbf{q}}''\right)
\frac{\left\langle {\mathbf{q}}+\frac{1}{2}{\mathbf{q}}'',{\mathbf{q}}''
|\hat{T}| {\mathbf{q}}_{0}\varphi_{d}\right\rangle}
{E-\frac{3}{4m}q''^{2}-E_{d}+i\varepsilon}.
\label{elastic-amp}
\end{eqnarray}
The amplitude for the full breakup process is given according to
Eq.~(\ref{eq:2.1c}) by
\begin{eqnarray}
\langle
{\mathbf{p}}{\mathbf{q}}|U_{0}|{\mathbf{q}}_{0}\varphi_{d}\rangle
&=&\frac{\left\langle
{\mathbf{p}}{\mathbf{q}}|\hat{T}|{\mathbf{q}}_{0}\varphi_{d}\right\rangle}
{E-\frac{3}{4m}{\mathbf{q}}^{2}-E_{d}} \cr
&+& \frac{\left\langle
-\frac{1}{2}{\mathbf{p}}+\frac{3}{4}{\mathbf{q}},-{\mathbf{p}}-\frac{1}{2}{\mathbf{q}}|\hat{T}|
{\mathbf{q}}_{0}\varphi_{d}\right\rangle}{E-\frac{3}{4m}(-{\mathbf{p}}-\frac{1}{2}
{\mathbf{q}})^{2}-E_{d}} + \frac{\left\langle
-\frac{1}{2}{\mathbf{p}}-\frac{3}{4}{\mathbf{q}},+{\mathbf{p}}-\frac{1}{2}{\mathbf{q}}|\hat{T}|
{\mathbf{q}}_{0}\varphi_{d}\right\rangle}
{E-\frac{3}{4m}(+{\mathbf{p}}-\frac{1}{2}{\mathbf{q}})^{2}-E_{d}} ~.
\nonumber \\
\label{breakup-amp}
\end{eqnarray}
Both operators are explicitly given in Ref.~\cite{Liu:2004tv} using
the independent variables of Eq.~(\ref{variables}), and can be 
directly applied with the expression of Eq.~(\ref{eq:4.9}).


\section{Summary and Conclusions}

In Ref.~\cite{Liu:2004tv} the formulation of the nonrelativistic
Faddeev equation for three
identical bosons as function of vector variables was introduced 
and successfully
solved for
laboratory projectile energies up to the GeV regime. The key point allowing the
calculation at those higher energies is to neglect the partial-wave
decomposition generally used at lower energies and to work directly with
momentum vectors, thus including all partial waves automatically. In the
formulation of the Faddeev integral equation in the continuum which is
most widely used, the singularities of the free three-body propagator occur 
simultaneously in an angle and momentum integration, leading to the
socalled logarithmic singularities. They require special care in 
numerical applications. Although sophisticated algorithms have 
been developed to integrate those singularities along the real axis, it
still is desirable to have a formulation of the Faddeev kernel, in which
the singularity structure is simpler. 

Starting from the formulation given in  Refs.~\cite{Liu:2004tv} 
and ~\cite{Witala:2008my}
 we propose
a new formulation of the Faddeev kernel which does not contain this
technical obstacle of logarithmic singularities. Instead, the singularities of
the free three-nucleon propagator appear as poles in a single variable. 
Those kind of poles can be integrated with standard techniques as e.g. 
used in integrating the two-body Lippmann-Schwinger equation. 
The singularity given by the deuteron pole is a simple pole (as before), but is now
cleanly separated in a separate integration. The integration over the angle
variable not affected by the pole structure remains very similar. 
This simplification in handling the singularities of the three-body 
continuum in a similar fashion as the two-body continuum should ease
applications of the Faddeev integral equations in different areas of physics.

\section*{Acknowledgments}
This work was performed in part under the
auspices of the U.~S.  Department of Energy, Office of
Nuclear Physics under contract No. DE-FG02-93ER40756 
with Ohio University and in part through JUSTIPEN (Japan-U.S. Theory
Institute for Physics with Exotic Nuclei) under grant NO.
DEFG02-06ER41407 with the University of Tennessee. It was also  partially 
supported by the Helmholtz Association
through funds provided to the virtual institute ``Spin and strong
QCD''(VH-VI-231) and by the  Polish Committee for Scientific
Research. Ch.E. thanks JUSTIPEN
for the warm hospitality during her stay at RIKEN.



\clearpage

\noindent

\end{document}